# 2D implementation of quantum annealing algorisms for fourth order binary optimization problems

Yong-Chao Tang,[1,*]  Guo-Xing Miao[1]

[1] Institute for Quantum Computing and Department of Electrical and Computer Engineering,
University of Waterloo, 200 University Ave W, Waterloo, ON, Canada, N2L 3G1

[*] Corresponding author. E-mail: yongchao.tang@uwaterloo.ca.

**Quantum annealing may provide advantages over simulated annealing on solving some problems such as $K^{th}$ order binary optimization problem. No feasible architecture exists to implement the high-order optimization problem ($K > 2$) on current quantum annealing hardware. We propose a two-dimensional quantum annealing architecture to solve the 4$^{th}$ order binary optimization problem by encoding four-qubit interactions within the coupled local fields acting on a set of physical qubits. All possible four-body coupling terms for an *N*-qubit system can be implemented through this architecture and are readily realizable with the existing superconducting circuit technologies. The overhead of the physical qubits is O($N^4$), which is the same as previously proposed architectures in four-dimensional space. The equivalence between the optimization problem Hamiltonian and the executable Hamiltonian is ensured by a gauge invariant subspace of the experimental system. A scheme to realize local gauge constraint by single ancillary qubit is proposed.**

**Introduction**

Quantum annealing (QA), being the quantum version of the widely used simulated annealing (SA) optimization method, has attracted substantial academic and industrial interest and inspired a rich body of literature on QA theories (1-5), applications (6-10), and experiments (11-14). The general paradigm of quantum annealing is to encode an optimization problem onto an objective Hamiltonian function of the *K*-spin model $H_P(\sigma)$. Its general form is

$$H_P(\sigma) = -\sum_{k=1}^{K} \sum_{j_1,\cdots,j_k=1}^{N} J_{j_1,\cdots,j_k} \sigma_z^{(j_1)} \cdots \sigma_z^{(j_k)},  \quad (1)$$

where $N$ is the problem size, $\sigma_z^{(j)}$ is the Pauli-Z matrix associated with the $j_{th}$ spin and the couplings $J_{j_1,\cdots,j_k}$ are real scalars. The task of finding the optimal solution to this optimization problem is converted into finding the ground state of $H_P$. If the evolution is sufficiently slow, quantum annealing can smoothly transfer an experimental qubits system from a trivial initial state, for example, the ground state of $H_I = \sum_{j=1}^{N} \sigma_x^{(j)}$, to the ground state of the objective Hamiltonian $H_P$. The whole time-dependent Hamiltonian of the system is

$$H(t) = \alpha(t/T)H_I + \beta(t/T)H_P \quad (0 \leq t \leq T), \tag{2}$$

where t is the time, T is the total time of the sweep, and $\alpha(t/T)$, $\beta(t/T)$ could be any functions with $\alpha(0)=1, \beta(0)=0$, and $\alpha(1)=0, \beta(1)=1$. Under adiabatic evolution, a programmable quantum annealer eventually reaches the ground state of $H_P$ with the aid of quantum tunneling, and one has thus found the desired result of the optimization problem.

Quantum tunneling can help QA penetrate high and narrow barriers which SA can hardly overcome, thereby exponential speedup may be achieved for certain types of problems. Among the numerous hard problems, $K^{th}$ order binary optimization problem with $K > 2$ is perhaps the best example problem to which quantum annealing offers a runtime advantage (15). $K^{th}$ order optimization problem is NP-hard and has many real-world applications in engineering and computational tasks. Its energy landscape gets more rugged with higher $K$. Since QA has advantages over SA with quantum tunneling on a rugged landscape, there may be larger subsets of instances for which QA runs faster as the order K increases.

One of the key challenges to map a $K^{th}$ order optimization problem onto an analog quantum annealer is to realize the *K*-body coupling terms. The quantum annealing hardware to-date is built from superconducting qubits (16) and couplers (17, 18), and only supports pairwise qubit couplings (*K* = 2). Therefore, it is difficult to lay out *K* local couplers on a two-dimensional chip or in a three-dimensional architecture for *K* > 2. Lechner *et al.* (19) have proposed a scalable architecture with all-to-all connectivity for two-qubit interactions. Although it can be extended to four-body and higher-order *K*-body interaction terms, it is impossible to arrange the constraints in three-dimensional space. Even for three-qubit interaction terms, the qubits have to be arranged in an infinite three-dimensional architecture. The superconducting researchers usually try to avoid three-dimensional implementations because there are many problems that need to be addressed, such as the placement of measurement circuitry etc., let alone the requirement for infinite space along z axis.

In this paper, we propose a two-dimensional architecture implementing a Hamiltonian for the 4$^{th}$ order optimization problem including four-qubit coupling terms formed with only local pairwise couplings. All possible four-body interactions for an *N*-qubit system can be encoded into this architecture and are realizable with existing programmable quantum annealers. This architecture may provide an opportunity to study open challenges in quantum annealing such as the role of the two-dimensional nature of the plaquette constraints during the sweep, and the scaling of quantum fluctuations on an existing programmable quantum annealer. Detailed implementation of this architecture is introduced in the next section, followed by the comparison between the energy spectrum of the problem Hamiltonian and the executable Hamiltonian.

**Implementation Details**

The 4$^{th}$ order binary optimization problem can be mapped onto an objective Hamiltonian of a four-spin model consisting of arbitrary four-qubit interaction terms $H_P = -\sum_{j_1,\cdots,j_4=1}^{N} J_{j_1 j_2 j_3 j_4} \sigma_z^{(j_1)} \sigma_z^{(j_2)} \sigma_z^{(j_3)} \sigma_z^{(j_4)}$, where $\sigma_z^{(j)}$ is the Pauli-Z matrix of $j_{th}$ qubit. The solution to the optimization problem is then transformed to finding the ground state of the objective Hamiltonian. Since their off-diagonal matrix elements in the standard product basis are all zero, it is easy to find that the $i_{th}$ eigenstate $|\psi_i\rangle$ to this kind of stoquastic Hamiltonian is a separable state. It takes the form $|\psi_i\rangle = |\phi_1\rangle \otimes \cdots \otimes |\phi_j\rangle \otimes \cdots \otimes |\phi_N\rangle$, and $|\phi_j\rangle$ is one of the two computational bases of single qubit. Thus, $|\psi_i\rangle$ is also the eigenstate to every four-qubit coupling term $h_P = -J\sigma_z^{(j_1)} \sigma_z^{(j_2)} \sigma_z^{(j_3)} \sigma_z^{(j_4)}$. We have $h_P |\psi_i\rangle = \pm J |\psi_i\rangle$. The main insight is that the contribution of a four-qubit coupling term to the energy of the whole problem Hamiltonian is determined by the product of the eigenvalues to the four single-qubit states in the computational basis $\sigma_z^{(j)} |\phi_j\rangle = \lambda_j |\phi_j\rangle$, where $\lambda_j = \pm 1$. The energy change applied by each coupling term to the whole system is thus $+J$ if $\lambda_{j_1} \lambda_{j_2} \lambda_{j_3} \lambda_{j_4}$ contains odd numbers of -1; otherwise, the energy change should be $-J$.

Inspired by Lechner *et al.* (19), we use $M = N(N+1)/2$ physical qubits from gauge constraints to represent N logical qubits and any pair of them in the problem Hamiltonian, where the eigenvalue of each physical qubit encodes the eigenvalues of $\sigma_z^{(j_1)} \sigma_z^{(j_2)}$ in the computational bases. Thus, a four-qubit interaction term $h_P$ is encoded with a two-qubit coupling term $\tilde{h}_P = -J\tilde{\sigma}_z^{(j_1,j_2)} \tilde{\sigma}_z^{(j_3,j_4)}$. When one of the two physical qubits represents a single logical qubit, we can also encode a three-qubit interaction term into a two-qubit coupling term. All possible three- and four-qubit interactions for an *N*-qubit system can be mapped onto two-qubit interactions and realized with existing quantum circuit technologies. The Hamiltonian of the four-spin model $H_P$ is encoded in the executable Hamiltonian

$$\tilde{H}_P = -\sum_{j_1,\cdots,j_4=1}^{N} J_{j_1 j_2 j_3 j_4} \tilde{\sigma}_z^{(j_1,j_2)} \tilde{\sigma}_z^{(j_3,j_4)} + \sum_{l=1}^{M-N} C_l. \tag{3}$$

where *M* is the number of physical qubits and $C_l$ is the local constraint that keeps the energy spectrum of the executable Hamiltonian consistent with the original problem Hamiltonian.

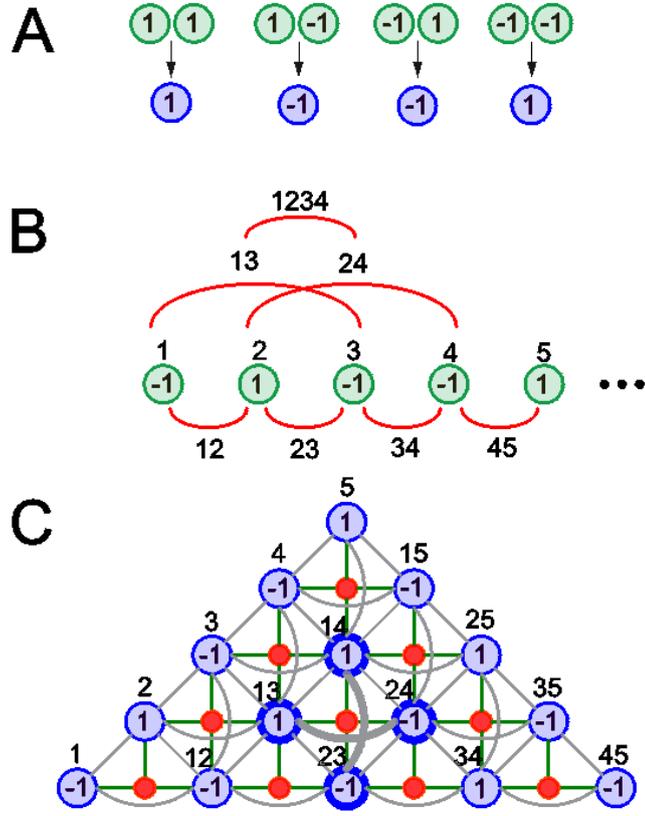

**Fig. 1. Illustration of the two-dimensional architecture. (A)** New encoding qubit variables are introduced for each of the *N* logical qubits and each of the *N*(*N*-1)/2 interactions, which take the value 1 if the two connected logical spins point in the same direction and -1 otherwise. **(B)** The aim is to encode a system of *N* logical spins consisting of three- and four-qubit interactions. A part of the interactions are shown with solid lines. **(C)** The two-dimensional architecture corresponds to five logical qubits consisting of four-qubit interactions. The blue circles are physical qubits, red dots are ancillary qubits for each plaquette. Solid bonds between two qubits are the pairwise couplings to realize the local constraints. The bold solid bond between qubits 14, 23 represents the sum of the couplings for the local constraints and for a four-qubit interaction term $\sigma_z^{(1)}\sigma_z^{(2)}\sigma_z^{(3)}\sigma_z^{(4)}$. The four-qubit interaction term can also be realized by the coupling between qubits 13, 24.

**Local constraints**

The successful transformation from four-qubit coupling terms into two-qubit interactions relies on the consistency of the spin values among all physical qubits. We encode two connected spins pointing in the same direction as 1, and -1 otherwise (Fig.

1A). The spin direction of a single qubit is also encoded as $\pm 1$. Thus the relative spin directions of a pair of encoding qubits should be equal to any other ways of pairing the same four logical qubits. For example, the relative alignment of $\sigma_z^{(1,2)}\sigma_z^{(3,4)}$ should be equal to $\sigma_z^{(1,3)}\sigma_z^{(2,4)}$ and $\sigma_z^{(1,4)}\sigma_z^{(2,3)}$. This constraint should cover all encoding qubits and demands either none, two, or all four encoding qubits in a closed loop to be antiparallel. That is, the number of -1's in the four encoding qubits $\widetilde{\sigma}_z^{(1,3)}$, $\widetilde{\sigma}_z^{(2,3)}$, $\widetilde{\sigma}_z^{(1,4)}$, $\widetilde{\sigma}_z^{(2,4)}$ has to be even (bold circles in Fig. 1C).

The consistency is achieved by introducing a configuration space in which a set of consistent qubit states are specified. Similar state subspace is also relevant in the context of the gauge-invariant subspace in lattice gauge theories (20). The subspace is generated by applying local constraints on each plaquette of four qubits in a two-dimensional array, which is an individual tile enclosed by a set of edges - here a square of four physical qubits.

In order to realize the local constraints with existing superconducting qubits and couplers, we propose to use an ancillary qubit for each plaquette in the qubits array. The form of the constraint is

$$C_l = +C(\sum_{m=w,n,e} \widetilde{\sigma}_z^{(l,m)} - \widetilde{\sigma}_z^{(l,s)} - 2\sigma_z^{(l)})^2, \qquad (4)$$

where $C$ denotes the energy scale for the constraint term, and the letters w, n, and e represent the qubits on the west, north, east. The sum runs over any three members of each plaquette, and then subtracts the spin value of the last one. $\sigma_z^{(l)}$ is an ancillary qubit for the $l_{th}$ constraint.

Next, the boundaries of the architecture have to be taken care of. We introduce a separate constraint on the hypotenuse of the triangle array (Fig. 1C) which consists of triangles instead of squares. The constraint enforces the condition that the number of 1's in each of these triangles is odd and its form is

$$C_l = +C(\sum_{m=w,n,e} \widetilde{\sigma}_z^{(l,m)} - 2\sigma_z^{(l)} - \mathrm{I})^2, \qquad (5)$$

The sum runs over all three members of each triangle, and $\mathrm{I}$ is an identity matrix. Only one ancillary qubit is used. Note that the constraint also involves local fields of physical qubits in each triangle. These two forms of constraints can be readily implemented by existing pairwise couplers and superconducting qubits.

**Implement the required four-qubit interactions**

As a final step, encoding all possible four-qubit interactions with two-qubit coupling terms in the physical qubits array needs long-range interactions that cannot be realized with the current technology. For the case of five logical qubits, to implement one of the four-qubit interaction terms $\sigma_z^{(1)}\sigma_z^{(2)}\sigma_z^{(3)}\sigma_z^{(5)}$ requires a coupling such as $\widetilde{\sigma}_z^{(1,2)}\widetilde{\sigma}_z^{(3,5)}$. The needed physical qubits are not neighboring to each other (Fig. 1C). To realize the four-qubit interactions required by the optimization problem, we have two proposals to overcome the challenge.

First, since both four-qubit interactions and all local constraints are implemented by pairwise couplings, we can treat the resultant one as a new problem Hamiltonian and encode the physical qubits with a set of new physical qubits. With the same architecture, a second-time encoding can allow all four-qubit couplings required in the original problem Hamiltonian to be realized by the local fields of new physical qubits. The overhead of physical qubits is O($N^4$), which is on the same order as the previously proposed architecture in four-dimensional space (19). Every logical qubit and every two-qubit interaction term is encoded $N$ times, while every three- or four-qubit interaction term is encoded 3 times. Although this kind of repetition coding alleviates the error rate and makes the annealing process more robust against spin flips from decoherence, there are also a large portion of physical qubits which not only encode the ancillary qubits in the first-time encoding architecture but also all four-qubit interactions repeatedly, which decreases the minimal energy gap on the spectrum of the final Hamiltonian. Furthermore, The O($N^4$) overhead of physical qubits places a limit on the problem scale of the real applications which can be solved.

Second, many real optimization problems will necessitate only L=O($N$) coupling terms, thus not all possible high-order interactions are needed. We can split the final encoding Hamiltonian into several executable parts, and implement each part with a triangle array and concatenate them along their catheti with ferromagnetic couplings. For example, all possible four-qubit interactions in a Hamiltonian of five qubits can be split into two part: $H_{p1} = -J_1 \sigma_z^1 \sigma_z^2 \sigma_z^3 \sigma_z^4 - J_2 \sigma_z^1 \sigma_z^2 \sigma_z^4 \sigma_z^5 - J_3 \sigma_z^2 \sigma_z^3 \sigma_z^4 \sigma_z^5$ and $H_{p2} = -J_4 \sigma_z^1 \sigma_z^2 \sigma_z^3 \sigma_z^5 - J_5 \sigma_z^1 \sigma_z^3 \sigma_z^4 \sigma_z^5$. Each part is embedded into a triangle array, and two arrays are combined together with ferromagnetic couplings (Fig. 2). The permutation of physical qubits can transform a long-range interaction of one array into a nearest-neighbor interaction in the other. Each triangle qubit array can be coupled on either of its catheti because either chain of qubits is a determining solution to the objective Hamiltonian. Thus this architecture is suitable for any required four-qubit couplings.

This architecture can be scaled up to $K^{th}$ order multi-qubit interactions for $K > 4$ and $K$ is even. The overall overhead of physical qubits is O($N^K$). With increasing order of the coupling terms, the requirement for implementation precision surges because the energy scale of local constraints will dominate the problem Hamiltonian. Both the qubits cost and the demanding constraints energy scale pose challenges for the required programmable quantum annealer. However, future technology development on qubit and coupler designs and on superconducting materials can eventually overcome these obstacles.

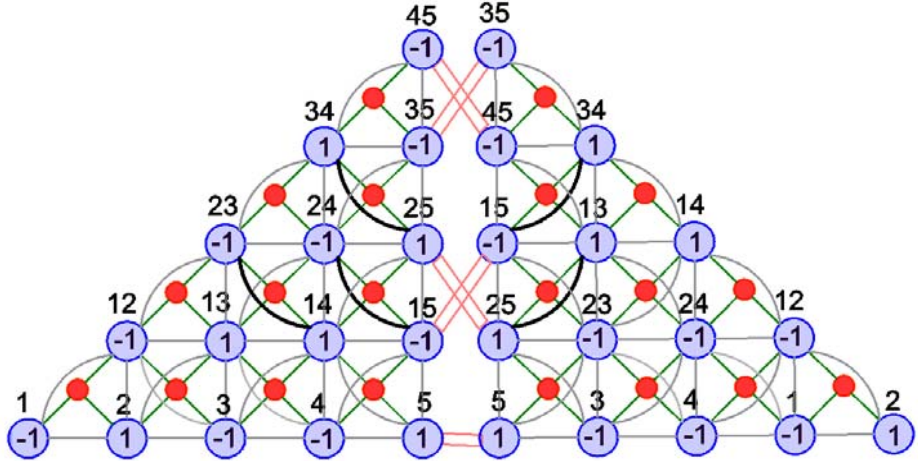

**Fig. 2. Scalable scheme.** The scheme can be scaled to realize any four-qubit interactions in a 4$^{th}$ order binary optimization problem. The problem consists of five qubits. The big blue circles represent the physical qubits, the small red circles denote ancillary qubits in each plaquette. The required four-qubit interactions are implemented by bold bonds between physical qubits. The two triangle arrays are coupled through ferromagnetic couplings (double lines) among physical qubits on one of the two catheti.

**Readouts**

The two schemes need different ways of readouts. For the first scheme, an appropriate way of decoding the encoded results is to make use of repetition codes and a majority-vote method. For simplicity, we focus on decoding the physical qubits representing single logical qubits and two-qubit interactions. Both are repeatedly encoded $N$ times. First, we decode spin values of single logical qubit and state configurations of two-qubit interactions out of the N-repetition codes. We can have robust readouts by decoding each N-repetition code. Second, we make use of a majority-vote method to obtain the final spin configuration for the original problem Hamiltonian.

After the first step of decoding, there are *N*(*N*+1)/2 spin values . The solution of the optimization can be fully determined by reading out an adequate choice of *N* among the *N*(*N*+1)/2 spin values. For example, in the case of *N*=5, one of the straightforward choices of readouts is the spin values for each logical qubit 1, 2, 3, 4, and 5. Other chains of relative configurations of pairs such as 5, 15, 25, 35, and 45, also hold the same information. A majority-vote method can be used among all possible combinations of *N*(*N*+1)/2 spin values to improve the accuracy. For the second scheme, each triangle array holds *N*(*N*+1)/2 spin values. We can get the readout of each triangle block and apply a majority-vote method afterwards.

# Fault tolerance

The 2D architecture can be interpreted as a classical error-correcting code because of its redundant encoding of the logical qubits information in the physical qubits (21). Thus it is highly robust against weakly correlated bit-flip noise. We use the product of $P_aP_b$ to determine the probability of retrieving errors from reading out all physical qubits when some spins are flipped as a result of decoherence. $P_a$ is the probability that an erroneous readout of a spin value is obtained after the adiabatic sweep. $P_b$ is the probability that a measurement indicates an erroneous solution due to this error.

For the first scheme, we do not take into account the ancillary qubits used in the first-time encoding because their readouts do not affect the solution. With N-repetition codes, the error rates for the readouts of single logical qubit and two-qubit interaction terms are $P_r = \sum_{i=\lceil N/2 \rceil}^{N} \binom{N}{i} P_e^i (1-P_e)^{N-i}$, where $P_e = \Gamma T$ is the probability for a spin flip of a physical qubit due to decoherence, and $\Gamma$ is the decoherence rate, and T is the total time of the adiabatic sweep. We have $N(N+1)/2$ robust readouts and get $P_a = P_r N(N+1)/2$. The information content of a single readout of a spin value is given by the ratio between the determining solutions that contain the given readout, $N_g$, and the total number of possible determining solutions $N_{all}$, $P_b = N_g/N_{all} = 2/N$. The product $P_aP_b = (N+1)P_r$ is the probability for errors (dashed lines in Fig. 3). A majority-vote method on the $N(N+1)/2$ robust readouts will give the correct answer as long as less than $N/4$ readouts from the N-repetition codes are faulty.

In the second scheme, each triangle array can be decoded independently. Suppose there are $M$ triangle arrays combined, then $P_a = \Gamma T M N(N+1)/2$ and $P_b = 2/MN$. Thus the error rate scales linearly with the problem size N: $P_aP_b = (N+1)\Gamma T$ (solid lines in Fig. 3). Applying a majority-vote method on all $MN(N+1)/2$ readouts can give the correct answer as long as less than $MN/4$ readouts are compromised.

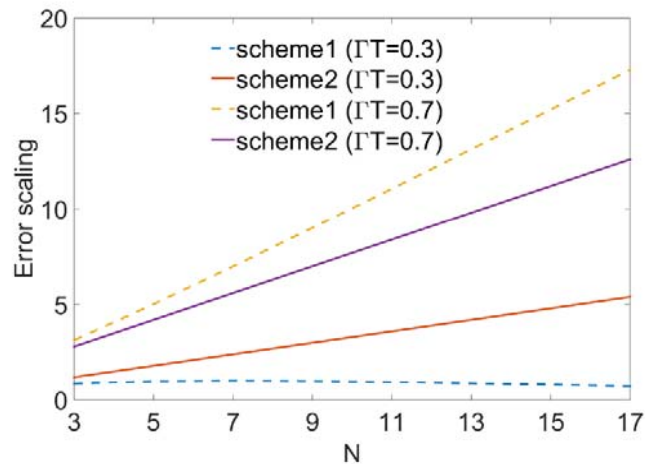

**Fig. 3. Error tolerance.** For the first scheme, the total error depends on $P_e = \Gamma T$, the probability for a spin flip due to decoherence. When $P_e < 0.5$, the N-repetition codes for single logical qubit and two-qubit interactions decrease the error of the final decoding and the total error decreases with increasing N (dashed blue). When $P_e \geq 0.5$, the total error rises with increasing N (dashed orange). For the second scheme, the total error in both schemes scale linearly with N (solid red and solid purple).

**Energy spectrum**

The protocol to find the ground state of an executable Hamiltonian is the same as in the original K-spin quantum annealing described in Eq. 2. We choose the ground state of a simple Hamiltonian that can be adiabatically transformed into Eq. 3. The simplest form for illustration could be

$$\widetilde{H}_I = \sum_{m=1}^{M} h_m \widetilde{\sigma}_x^{(m)}, \tag{6}$$

where $M = N^2$ is the number of all qubits including the ancillary qubits for the constraints. The adiabatic sweep is described by the time-dependent Hamiltonian

$$\widetilde{H}_{sweep}(t) = \alpha(t/T)\widetilde{H}_I + \beta(t/T)\widetilde{H}_P \quad (0 \leq t \leq T), \tag{7}$$

In this architecture, the constraints should be modulated by $\beta(t/T)$, otherwise the low-energy levels are so closed to each other that the minimal gap is almost zero at the beginning of the sweep. In the second scheme of combining several triangle arrays, the three- and four-qubit interactions are realized through the couplers which also implement the couplings for the constraints. Those couplings will be dominated by the energy terms from the constraints. Usually the couplings have to be normalized by a factor to within the range [-1,1], which is the operational range in current annealer chips. Thus the couplers must be able to achieve high precision when implementing the interaction terms.

The new time-dependent Hamiltonian is embedded in a larger Hilbert space and restricted in a gauge invariant subspace by the local constraints. It has a different spectrum and the sweep is associated with a different quantum path compared to the adiabatic optimization in Eq. 2. The difference between the two sweeps is illustrated in Fig. 4. The lowest states in the final Hamiltonian are identical in both representations of the optimization problem. However, the minimal gap in the direct implementation using interacting logical qubits is larger than the minimal gap in the executable architecture, and the ratio between these gaps increases with the energy scale C/J, see Fig. 5. The calculated gap ratio is larger than what is obtained in (19) because we use ancillary qubits instead of four-qubit interactions to realize the local constraints.

Instead of calculating the energy levels of the second scheme, we compare the spectrum of the executable architecture implemented by single triangle array with an ideal sweep realized by logical qubits. Due to the limited technical progress on superconducting couplers at the moment, no other realistic way to tackle a Hamiltonian with high-order interactions on a large scale has been proposed, so this comparison should only serve as a benchmark. The second scheme uses more physical qubits, but it is supposed to be applied to sparse high-order interactions. In this scenario, only several times more qubits are consumed. Compared with the architecture consisting of a single triangle array, the second scheme should lead to smaller but comparable energy gaps.

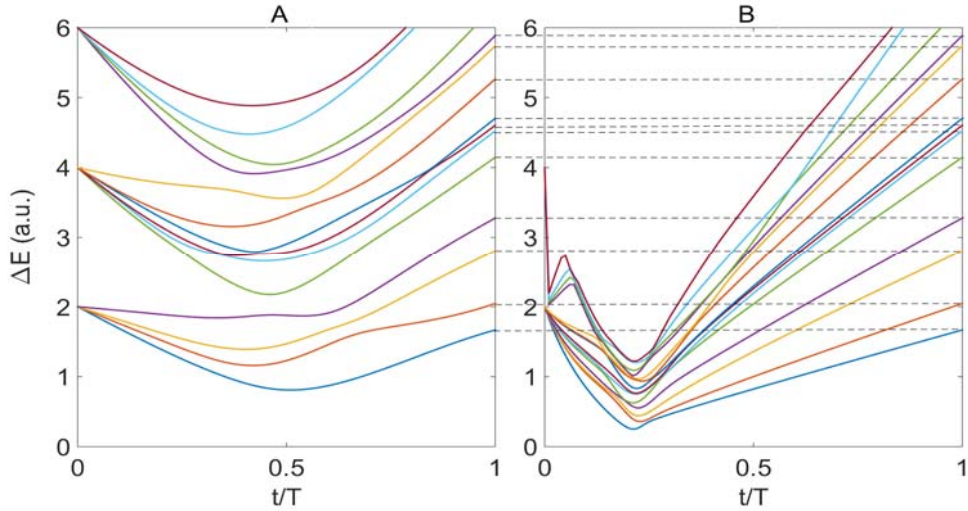

**Fig. 4. Time-dependent spectrum. (A)** Energy spectrum of a typical adiabatic passage with N=4 logical qubits with four-qubit interactions in a fictitious implementation of the logical qubits. **(B)** The evolution of the executable Hamiltonian for the same problem implemented with ancillary qubits. Here, t is the time and T is the total time of the sweep. Instantaneous eigenenergy $E_i$ is measured with respect to the ground state, $\Delta E = E_i - E_0$. The coupling strengths for two-, three-, and four-qubit interactions and the local field of every logical qubit are random numbers uniformly taken from the interval [-J,J], and the constraint strength is C/J = 2. At the end of the evolution, an exact correspondence is achieved between the lowest levels of the two-dimensional architecture and the original model of classical spins (dashed lines).

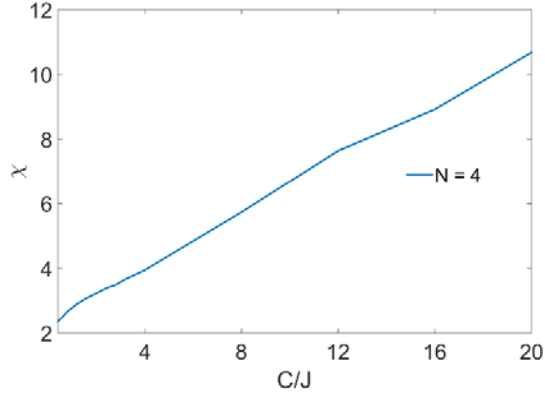

**Fig. 5. Ratio of energy gap.** The ratio $\chi$ of the minimal gap during the adiabatic optimization between the logical system and the executable architecture, as a function of C/J, for *N*=4.

**Optimal energy scale**

The role of the energy scale $\beta = C/J$ is important for the successful implementation of the quantum annealing procedure. When $\beta \ll 1$, the constraints have no effect. When $\beta \gg 1$ the constraint dominates the problem scale and the comparatively small interaction terms implemented by the couplers can be easily affected by noises. Thus, there should be an optimal $\beta$ for different problem Hamiltonian, which we denote as $\beta_{opt}$. We expect $\beta_{opt}$ to be around 2. To understand the role of $\beta_{opt}$, consider first how increasing $\beta$ affects the size and position of the gap $\Delta E$. The excitations relevant to our local constraints are from the first excited state and above. In Fig. 6A, we show that the relevant gap dwindles with increasing $\beta$ when $\beta \leq 2$, and remains unchanged when $\beta$ is increased further. The minimum gap position also develops over a shorter time, which is advantageous since it leaves less time for thermal excitations to act while the transverse field dominates. However, the role of $\beta$ is more subtle than suggested by considering only the gap. The energy states outside the gauge invariant subspace may also emerge in the low-energy spectrum. These additional energy levels are also determined by $\beta$. More impurity states may emerge in the low-energy spectrum with decreasing $\beta$. Fig. 6B shows the dependence of the lowest impurity state on $\beta$, which encourages us to choose a larger $\beta$. On the other hand, Fig. 5 shows the ratio of the minimal gap during the adiabatic optimization between the logical system and the executable architecture grows with increasing $\beta$. The real optimal value should be determined by experiments, with the above considerations in mind.

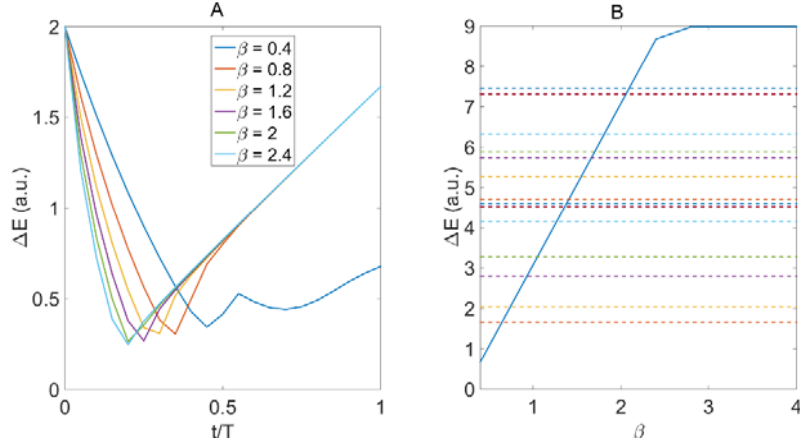

**Fig. 6. Effect of the energy scale of local constraints. (A)** The relative first excited state varies with increasing $\beta$. **(B)** The dependence of the lowest impurity state (induced by constraint terms) on $\beta$. The energy of the impurity state increases with increasing beta and surpasses all energy levels of the problem Hamiltonian (dashed lines) at $\beta = 2.2$

**Potential barrier for encoding Hamiltonian**

There is a concern on the role of ancillary qubits over the energy landscape of the Hamiltonian (15). In quantum annealing, spin tunneling can be described in mean field spin models using the path integral over spin-coherent states (22-24). The dynamics is dominated by the paths through the mean field energy landscape that have the highest transition probabilities. The mean-field potential is of the form

$$V(\hat{m}) = \langle \Psi_{\hat{m}} | \tilde{H}(t) | \Psi_{\hat{m}} \rangle, \qquad (8)$$

where $\tilde{H}(t)$ is the time-dependent QA Hamiltonian (6) and $|\Psi_{\hat{m}}\rangle$ is a product state

$$|\Psi_{\hat{m}}\rangle = \bigotimes_j [\cos\frac{\theta_j}{2}|0\rangle + e^{-i\phi_j}\sin\frac{\theta_j}{2}|1\rangle]. \qquad (9)$$

The coherent state of the $j$th spin is defined by a vector on the Bloch sphere

$$\mathbf{n}_j = (\sin\theta_j \cos\phi_j, \sin\theta_j \sin\phi_j, \cos\theta_j), \qquad (10)$$

and the corresponding state of the N-qubit system is defined by the tensor $\hat{\mathbf{m}} = (\mathbf{n}_1, \mathbf{n}_2, \cdots, \mathbf{n}_N)$.

The mean-field potential can be split into two parts $V_1(\hat{m}) = \langle \Psi_{\hat{m}} | \tilde{H}(t) - \beta(t)\sum_l C_l | \Psi_{\hat{m}} \rangle$ and $V_2(\hat{m}) = \beta(t)\sum_l \langle \Psi_{\hat{m}} | C_l | \Psi_{\hat{m}} \rangle$. Each ancillary qubit appears only in one term of $V_2(\hat{m})$. We rewrite each constraint term as $C_l = C_l^{(1)} + C_l^{(2)} + 8I$, where $C_l^{(1)} = -4C(\sum_{m=w,n,e}\tilde{\sigma}_z^{(l,m)} - \tilde{\sigma}_z^{(l,s)})\sigma_z^{(l)}$ is the part involving an

ancillary qubit, $C_l^{(2)}$ is the rest part of the constraint term. The form of $C_l^{(1)}$ is similar to the energy penalty terms used in the quantum annealing error correction codes (25, 26). Both the constraints and the energy penalty terms are to restrict the whole system into a specified code space and to punish the states outside this space.

The effect of the two-dimensional local constraints on the energy landscape of an *N*-qubit system is an open question. The width of a potential barrier is measured with the change of the state of an *N*-qubit system between two local minima. For a rugged energy landscape, a narrow barrier means small changes of the state of the *N*-qubit system. Since every ancillary qubit is only coupled with three- or four-nearest neighbor qubits, small changes of the *N*-qubit states may leave most of ancillary qubits unvaried.

**Conclusions**

In summary, we have presented an executable two-dimensional quantum annealing architecture for solving the 4$^{th}$ order binary optimization problem. It features all possible four-qubit interactions and O(*N*$^4$) overhead of physical qubits. It also can be scaled up to any four-qubit coupling terms without repeatedly encoding the same interactions. The optimization problem is encoded in local fields acting on the qubits and the interactions between them. This allows one to implement an objective Hamiltonian for a high-order binary optimization problem, whose solution is encoded in the topology of the physical qubits. This architecture may be used to solve a valuable class of NP-hard problems - the *K*$^{th}$ order binary optimization problems.

**Acknowledgements:**

This work is supported by NSERC Discovery grant RGPIN 418415-2012.